\documentclass[aps,twocolumn]{revtex4}

\begin{document}
\pagestyle{empty}
\title{Transverse Solutions of the Vector Laplace and Helmholtz
Equations \\ for Spherical Coordinates and Boundary Conditions
with Azimuthal Symmetry}
\author{Ernesto A. MATUTE}
\affiliation{Departamento de F\'{\i}sica, Universidad de Santiago
de Chile, Casilla 307 - Correo 2, Santiago, Chile} \maketitle

\begin{center}
\small {\bf ABSTRACT}
\end{center}

\noindent A scalar field method to obtain transverse solutions of
the vector Laplace and Helmholtz equations in spherical
coordinates for boundary-value problems with azimuthal symmetry is
described.  Neither scalar nor vector potentials are used.
Solutions are obtained by use of separation of variables instead
of dyadic Green's functions expanded in terms of vector spherical
harmonics.  Applications to the calculations of magnetic fields
from steady and oscillating localized current distributions are
presented.
\\


\renewcommand{\thesection}{\arabic{section}}

\section{Introduction}

\noindent
In many physical applications the vector Laplace equation
$\nabla^{2} \vec{F} = \vec{0}$ or the vector Helmholtz equation
$\nabla^{2} \vec{F} + k^{2} \vec{F}= \vec{0}$ for spherical coordinates
and boundary conditions has to be solved.  It is the case for instance
with the electric and magnetic fields wherever there is no free charge
and current distributions.  In spherical coordinates, however, the
Laplacian of the vector $\nabla^{2} \vec{F}$ leads to a set of three
simultaneous equations, each equation involving all three components
$F_{r}$, $F_{\theta}$, $F_{\varphi}$ [1].  This complication
is well known and general techniques for solving these equations have
been developed, based on a dyadic Green's function involving vector
spherical harmonics which transforms the boundary conditions and source
densities into the vector solution~[2].
\\

\noindent In a number of cases the vector field must also have
zero divergence: $\vec{\nabla}\cdot\vec{F}=0$.  This is so for
example for the magnetic field and in general for transverse
vector fields.  With this transverse condition it can be shown
that the above vector equations can be separated into an equation
for $F_{r}$ alone.  However, this is not true for the other two
scalar components of the vector field. Nevertheless, it is
possible to obtain a straightforward solution for $F_{\theta}$ in
a problem with azimuthal symmetry; here we are not concerned about
those symmetry cases where only $F_{\varphi}$ is nonzero because a
scalar technique of separation of variables is already known to
obtain the solution~[3].  Several problems of physical interest
can then be treated with the standard mathematical methods used to
solving scalar field equations, so avoiding the dyadic method~[4].
\\

\noindent
The purpose of this paper is to report the use of the technique to
compute the solution to instructive examples taken from the
electromagnetic theory, though there are different other places where
vector fields show up.  Specifically, we work out magnetic field
solutions from localized current distributions.  The main methods which
may be used for solving boundary-value problems in magnetism involve
the vector potential and the scalar potential~[5].  The new
technique we discuss here deals with the magnetic field itself, without
reference to any kind of potential.  The approach can also be used to
calculate electric fields from localized charge distributions since the
equations to be solved are the same.

\section{Transverse solutions of the vector Laplace equation}

\noindent
In order to calculate the components of the vector $\nabla^{2} \vec{F}$
along the axes of the spherical system of coordinates, the following
identity must be used:
\begin{equation}
\nabla^{2} \vec{F} = - \vec{\nabla} \times (\vec{\nabla} \times \vec{F})
+ \vec{\nabla} (\vec{\nabla} \cdot \vec{F}) .
\end{equation}
By combining with the transverse condition $\vec{\nabla} \cdot
\vec{F} = 0$ and requiring azimuthal symmetry, we find the
equation
\begin{eqnarray}
(\nabla^{2} \vec{F})_{r} & = & \frac{1}{r^{2}} \;
\frac{\partial^{2}}{\partial r^{2}} (r^{2} F_{r}) + \frac{1}{r^{2}
\sin \, \theta} \; \frac{\partial}{\partial \theta} (\sin \,
\theta \; \frac{\partial F_{r}}{\partial \theta}) \nonumber
\\ & & \nonumber \\ & = & 0
\label{eqr}
\end{eqnarray}
for the radial component of the vector Laplace equation
$\nabla^{2} \vec{F} = \vec{0}$.  To compute $F_{\theta}$ is
convenient to use the transverse condition:
\begin{eqnarray}
\vec{\nabla} \cdot \vec{F} & = & \frac{1}{r^{2}} \;
\frac{\partial}{\partial r} (r^{2} F_{r}) + \frac{1}{r \sin \,
\theta} \; \frac{\partial}{\partial \theta} (\sin \, \theta \;
F_{\theta}) \nonumber \\ & & \nonumber \\ & = & 0 , \label{div}
\end{eqnarray}
where azimuthal symmetry has been demanded.
\\

\noindent
Equation (\ref{eqr}) can be solved by the usual technique of separation
of variables.  If a product form for $F_{r}$ is assumed, then it can be
written
\begin{equation}
F_{r}(r,\theta) = \frac{u(r)}{r^2} \; P(\theta) .
\end{equation}
The following separate equations for $u(r)$ and $P(\theta)$ are obtained:
\begin{eqnarray}
& & \frac{d^{2}u}{dr^{2}} - \frac{n(n+1)}{r^{2}} \; u = 0 ,  \label{req}
\\ & & \nonumber \\
& & \frac{1}{\sin \, \theta} \; \frac{d}{d \theta} (\sin \, \theta
\; \frac{d P}{d \theta}) + n (n + 1) P = 0 , \label{Legendre}
\end{eqnarray}
where $n(n+1)$ is the real separation constant.  The solution of
Eq.~(\ref{req}) is
\begin{equation}
u(r) = a \, r^{n+1} + \frac{b}{r^{n}} .
\end{equation}
Equation (\ref{Legendre}) is the Legendre equation of order $n$ and the
only solution which is single valued, finite and continuous over the
whole interval corresponds to the Legendre polynomial
$P_{n}(\cos \, \theta)$, $n$ being a positive integer.  Thus the general
solution for $F_{r}$ is
\begin{equation}
F_{r}(r,\theta) = \sum_{n=0}^{\infty} \left( a_{n} r^{n-1} +
\frac{b_{n}}{r^{n+2}} \right) P_{n}(\cos \, \theta) .
\label{radial}
\end{equation}
The simplest way of treating Eq.~(\ref{div}) is to use the series
expansion
\begin{equation}
F_{\theta}(r,\theta) = \sum_{n=0}^{\infty} v_{n}(r) \;
\frac{d}{d\theta} P_{n}(\cos \, \theta) , \label{angular}
\end{equation}
where $v_{n}(r)$ are functions to be determined.  By replacing
Eqs.~(\ref{radial}) and (\ref{angular}) into Eq.~(\ref{div}), it
is found that
\begin{equation}
v_{n}(r) = \frac{a_{n}}{n} \; r^{n-1} - \frac{b_{n}}{n+1} \;
\frac{1}{r^{n+2}}
\label{v}
\end{equation}
for $n \geq 1$ with $a_{o} = 0$.  The coefficients $a_{n}$ and $b_{n}$
are to be determined from the boundary conditions.
\\

\noindent On the other hand, we note that the solution obtained
from Eqs.~(\ref{radial}), (\ref{angular}) and (\ref{v}) satisfies
the angular component of the Laplace equation for transverse
vector fields and azimuthal symmetry:
\begin{equation}
(\nabla^{2} \vec{F})_{\theta} =
\frac{1}{r} \; \frac{\partial^{2}}{\partial r^{2}} (r F_{\theta}) -
\frac{1}{r} \; \frac{\partial^2 F_{r}}{\partial r \partial \theta} = 0 ,
\label{teta}
\end{equation}
as expected by consistency.  This happens because Eqs.~(\ref{eqr})
and (\ref{div}) imply
\begin{equation}
\frac{\partial}{\partial r} (r F_{\theta}) -
\frac{\partial F_{r}}{\partial \theta} = 0 ,
\label{good}
\end{equation}
which leads to Eq.~(\ref{teta}); clearly, the solutions given in
Eqs.~(\ref{radial}), (\ref{angular}) and (\ref{v}) satisfy
Eq.~(\ref{good}).
\\

\noindent
In the case of the magnetic field, we recall that the boundary conditions
at a boundary separating two regions are
\begin{eqnarray}
& & \hat{n} \cdot (\vec{B}_{1} - \vec{B}_{2}) = 0, \nonumber
\\ & & \nonumber \\
& & \hat{n} \times (\vec{H}_{1} - \vec{H}_{2}) = \vec{J}_{S} ,
\label{boundary}
\end{eqnarray}
where $\vec{J}_{S}$ is the surface current density and the normal unit
vector $\hat{n}$ is drawn from the second region into the first one.
For a linear medium the constitutive relation $\vec{B}=\mu\vec{H}$ holds,
$\mu$ being the constant magnetic permeability.  To see how the technique
works out, we consider the simple, common textbook example of the
magnetic field produced by the rotation of a sphere of radius $a$,
uniformly charged with a charge $Q$, with constant angular velocity
$\omega$, which is usually solved with the vector potential
method~[2,5].  The surface current density at $r=a$ is
\begin{equation}
\vec{J}_{S}(a,\theta,\varphi) = \frac{\omega Q}{4 \pi a} \;
\sin \, \theta \; \hat{\varphi} ,
\end{equation}
where the $z$ axis has been chosen along the rotation axis.  The
problem reduces to using the series solutions (\ref{radial}),
(\ref{angular}) and (\ref{v}) of the Laplace equation in the
regions $r>a$ and $r<a$ to conform to the requirement that the
magnetic field must be finite at the origin, vanish at infinity
and satisfy the boundary conditions of Eq.~(\ref{boundary}) at $r
= a$.  Using $\vec{B}=\mu_{\circ}\vec{H}$, where $\mu_{\circ}$ is
the vacuum magnetic permeability, the resulting magnetic field can
be seen to be
\begin{equation}
\vec{H}(r,\theta) = \frac{\omega Q}{12 \pi} \;
\left\{ \begin{array}{l}
        \displaystyle \frac{a^2}{r^3} \; (3 \cos \, \theta \; \hat{r} -
        \hat{z}) \; ; \; r > a \\  \\
        \displaystyle \frac{2}{a} \; \hat{z} \; ; \; r < a
        \end{array}
\right.  .
\end{equation}
This solution describes a uniform magnetic field inside the sphere and a
dipole field outside with dipole moment $\vec{m} = \hat{z} \; \omega Q
a^{2} / 3$.
\\

\noindent Another example is that of the magnetic field from a
circular current loop of radius $a$ lying in the
\linebreak $x$-$y$ plane with its center at the origin and
carrying a current $I$.  In this case the surface current density
at $r=a$ is
\begin{equation}
\vec{J}_{S}(a,\theta,\varphi) = \frac{I}{a} \; \delta(\cos \, \theta) \;
\hat{\varphi} .
\end{equation}
The radial component of the magnetic field turns out
\newpage
\noindent
to be now
\begin{eqnarray}
H_{r}(r,\theta) & = & \frac{I}{2} \sum_{n=0}^{\infty}
\frac{(-1)^{n} (2n+1)!!}{2^{n} n!} \; P_{2n+1}(\cos \, \theta) \nonumber
\\ & & \nonumber \\ & & \times
\left\{ \begin{array}{l}
        \displaystyle \frac{a^{2n+2}}{r^{2n+3}} \; ; \; r > a \\  \\
        \displaystyle \frac{r^{2n}}{a^{2n+1}} \; ; \; r < a
        \end{array}
\right.  ,
\end{eqnarray}
while the angular component becomes
\begin{eqnarray}
H_{\theta}(r,\theta) & = & \frac{I}{4} \sum_{n=0}^{\infty}
\frac{(-1)^{n} (2n+1)!!}{2^{n} (n+1)!} \; P_{2n+1}^{1}(\cos \, \theta)
\nonumber \\ & & \nonumber \\ & & \times
\left\{ \begin{array}{l}
        - \displaystyle \frac{a^{2n+2}}{r^{2n+3}} \; ; \; r > a \\  \\
        \left( \displaystyle \frac{2n+2}{2n+1} \right)
        \displaystyle \frac{r^{2n}}{a^{2n+1}} \; ; \; r < a
        \end{array}
\right.  ,
\end{eqnarray}
where $P_{2n+1}^{1}(\cos \, \theta) = (d/d\theta) \, P_{2n+1}(\cos \,
\theta)$ is an associated Legendre polynomial.

\section{Transverse solutions of the vector Helmholtz equation}

\noindent
In the case of the Helmholtz equation for transversal vector fields the
radial equation becomes
\begin{eqnarray}
& & (\nabla^{2} \vec{F})_{r} + k^{2} F_{r} = \frac{1}{r^{2}} \;
\frac{\partial^{2}}{\partial r^{2}} (r^{2} F_{r}) \nonumber \\ & &
\nonumber \\ & & \hspace{0.4cm} + \frac{1}{r^{2} \sin \, \theta}
\; \frac{\partial}{\partial \theta} (\sin \, \theta \;
\frac{\partial F_{r}}{\partial \theta}) + k^{2} F_{r} = 0 .
\label{Hreq}
\end{eqnarray}
Equation (\ref{div}) still applies for problems possessing
azimuthal symmetry; it is used to obtain $F_{\theta}$ from
$F_{r}$.
\\

\noindent
Next, by putting
\begin{equation}
F_{r}(r,\theta) = \frac{j(r)}{r} \; P(\theta) ,
\end{equation}
we can separate Eq.~(\ref{Hreq}) into two equations:
\begin{equation}
\frac{d^{2}j}{dr^{2}} + \frac{2}{r} \; \frac{dj}{dr} + \left[ k^{2} -
\frac{n(n+1)}{r^{2}} \right] j = 0
\label{Bessel}
\end{equation}
for $j(r)$ and Eq.~(\ref{Legendre}) for $P(\theta)$, where
$n(n+1)$ is the separation constant.  Equation (\ref{Bessel}) is
the spherical Bessel equation of order $n$ in the variable $kr$.
Therefore, the general solution for $F_{r}$ is
\newpage
\noindent
\begin{equation}
F_{r}(r,\theta) = \sum_{n=0}^{\infty} \left[ c_{n}
\frac{j_{n}(kr)}{r} + d_{n} \frac{n_{n}(kr)}{r} \right] P_{n}(\cos
\, \theta) . \label{Hradial}
\end{equation}
Depending on boundary conditions, the spherical Hankel functions
$h_{n}^{(1,2)}$ instead of the spherical Bessel functions $j_{n}$,
$n_{n}$ may be used.  For $F_{\theta}$ we again assume
\begin{equation}
F_{\theta}(r,\theta) = \sum_{n=0}^{\infty} w_{n}(r) \;
\frac{d}{d\theta} P_{n}(\cos \, \theta) , \label{Hangular}
\end{equation}
where for $w_{n}$ we now obtain
\begin{eqnarray}
w_{n}(r) & = & \frac{c_{n}}{n(n+1) r} \; \frac{d\;}{dr} [r \, j_{n}(kr)]
\nonumber \\ & & \nonumber \\ & &
+ \frac{d_{n}}{n(n+1) r} \; \frac{d\;}{dr} [r \, n_{n}(kr)] .
\label{w}
\end{eqnarray}
The constants $c_{n}$ and $d_{n}$ are computed from the requirement that
the vector field has to satisfy the boundary conditions of the given
problem.
\\

\indent On the other hand, we observe that the solution given in
Eqs.~(\ref{Hradial}), (\ref{Hangular}) and (\ref{w}) satisfies the
angular component of the Helmholtz equation for transverse vector
fields and azimuthal symmetry:
\begin{eqnarray}
(\nabla^{2} \vec{F})_{\theta} + k^{2} F_{\theta} & = &
\frac{1}{r} \; \frac{\partial^{2}}{\partial r^{2}} (r F_{\theta}) -
\frac{1}{r} \; \frac{\partial^2 F_{r}}{\partial r \partial \theta} +
k^{2} F_{\theta} \nonumber \\ & & \nonumber \\ & = & 0 ,
\end{eqnarray}
as expected, where Eq.~(\ref{Bessel}) must be used.
\\

\noindent
An illustration is provided by the problem of the radiation from a
vibrating current loop lying in the $x$-$y$ plane.  The surface density
current at $r=a$ is
\begin{equation}
\vec{J}_{S}(a,\theta,\varphi,t) = \frac{I_{\circ}}{a} \;
\delta(\cos \, \theta) \; e^{-i \omega t} \; \hat{\varphi}  .
\end{equation}
The series solution of the Helmholtz equation for the magnetic
field, which is finite at the origin, represents outgoing waves at
infinity and conforms to the boundary conditions of
Eq.~(\ref{boundary}) at $r=a$, is then given by
\begin{eqnarray}
& & \displaystyle H_{r}(r,\theta,t) = i \frac{I_{\circ} k a}{2 r}
e^{-i \omega t} \sum_{n=0}^{\infty}
\frac{(-1)^{n}(4n+3)}{2^{n}} \nonumber \\ & & \nonumber \\
& & \displaystyle \times \frac{(2n+1)!!}{n!} P_{2n+1}(\cos \, \theta)
\left\{ \begin{array}{l}
        \displaystyle j_{2n+1}(ka) \; h_{2n+1}^{(1)}(kr)
        \\  \\  \\
        \displaystyle j_{2n+1}(kr) \; h_{2n+1}^{(1)}(ka)
        \end{array}
\right.
\end{eqnarray}
and
\begin{eqnarray}
& & H_{\theta}(r,\theta,t) = i \frac{I_{\circ} k^{2} a}{2}
e^{-i \omega t} \sum_{n=0}^{\infty}
\frac{(-1)^{n}(4n+3)}{2^{n+1}(2n+1)} \nonumber \\ & &
\times \; \frac{(2n+1)!!}{(n+1)!} \; P_{2n+1}^{1}(\cos \, \theta)
\nonumber \\ & & \nonumber \\ & & \times
\left\{ \begin{array}{l}
        j_{2n+1}(ka) \left[ h_{2n}^{(1)}(kr) - \displaystyle
        \frac{2n+1}{kr} \; h_{2n+1}^{(1)}(kr) \right]
        \\  \\  \\
        h_{2n+1}^{(1)}(ka) \left[ j_{2n}(kr) - \displaystyle
        \frac{2n+1}{kr} \; j_{2n+1}(kr) \right]
        \end{array}
\right.
\end{eqnarray}
where the upper lines hold for $r>a$ and the lower lines for $r<a$.
The radiative part of the magnetic field takes on the limiting form
\begin{eqnarray}
\vec{H}(r,\theta,t) & = & \hat{\theta} \; \frac{I_{\circ} k a}{4 r}
e^{i(k r - \omega t)} \sum_{n=0}^{\infty}
\frac{(4n+3)(2n+1)!!}{2^{n}(2n+1)(n+1)!} \nonumber \\ & & \nonumber \\
& & \times \; j_{2n+1}(ka) \; P_{2n+1}^{1}(\cos \, \theta) .
\end{eqnarray}
The resulting magnetic field can be seen to be the same as that obtained
by another more extensive methods~[2].

\section{Conclusion}

\noindent The expressions for the components of $\nabla^{2}
\vec{F}$ in spherical coordinates are really messy, but they
appear for instance in Laplace and Helmholtz equations for vector
fields, such as the electromagnetic fields.  We have presented a
fairly simple technique of separation of variables to compute
transverse solutions with $\vec{\nabla} \cdot \vec{F} = 0$ of
these vector equations applied to boundary-value problems with
azimuthal symmetry, so avoiding the general, more extensive method
in which dyadic Green's functions and series expansions in vector
spherical harmonics are introduced.  It involves scalar solutions
for the spherical components of $\vec{F}$ which resemble the ones
obtained for the scalar Laplace and Helmholtz equations.  No
scalar or vector potentials are used.  Illustrations were provided
by problems concerned with the magnetic field from steady and
oscillating localized currents.  We finally remark that the method
can also be extended to solving boundary-value problems in
cylindrical coordinates possessing azimuthal symmetry, although in
this case we deal with a separated equation for each scalar
component of the vector field~[1], so that the problem of solving
the vector Laplace or Helmholtz equation reduces to that of
solving the corresponding scalar equation.
\\  \\

\section{Acknowledgments}
\noindent
This work was partially supported by the Departamento de Investigaciones
Cient\'{\i}ficas y Tecnol\'ogicas, Universidad de Santiago de Chile.

\section{References}

\newcounter{refe}
\begin{list}
{[\arabic{refe}]}{\usecounter{refe} \setlength{\leftmargin}{6mm}
\setlength{\parsep}{-1mm}}
\item G. Arfken and H. Weber, {\bf Mathematical Methods for
Physicists}, New York: Academic Press Inc., 2001, 5th edition, Chap. 2.
\item P.M. Morse and H. Feshbach, {\bf Methods of Theoretical Physics},
New York: McGraw-Hill Book Company Inc., 1953, Vol. 2, Chap. 13.
\item Reference [1], Chap. 12.
\item E.A. Matute, ``On the Superconducting Sphere in an External
Magnetic Field'', {\bf American Journal of  Physics}, Vol. 67, No. 9,
1999, pp. 786-788.
\item J.D. Jackson, {\bf Classical Electrodynamics},
New York: John Wiley \& Sons Inc., 1998, 3rd edition, Chap. 5.
\end{list}
\end{document}